# q-Dependence of the giant bond-stretching phonon anomaly in the stripe compound La$_{1.48}$Nd$_{0.4}$Sr$_{0.12}$CuO$_4$ measured by IXS

D. Reznik[1], T. Fukuda[2], D. Lamago[1,3], A.Q.R. Baron[4,5], S. Tsutsui[5], M. Fujita[6], and K. Yamada[6]

1. Forschungszentrum Karlsruhe, Institut für Festkörperphysik, P.O.B. 3640, D-76021 Karlsruhe, Germany

2. Synchrotron Radiation Research Unit, Japan Atomic Energy Agency (SPring-8), Sayo, Hyogo 679-5148, Japan

3. Laboratoire Léon Brillouin, CE Saclay, F-91191 Gif-sur-Yvette Cedex, France

4. Materials Dynamics Laboratory, Harima RIKEN, 1-1-1 Kouto, Sayo, Hyogo, 679-5148, JAPAN

5. Japan Synchrotron Radiation Research Institute, 1-1-1 Kouto, Sayo, Hyogo, 679-5198, JAPAN

6. Institute for Material Research, Tohoku University, Katahira, Aoba-ku, Sendai, 980-8577, Japan.

*abstract:* Inelastic x-ray scattering (IXS) was used to study the Cu-O bond-stretching vibrations in the static stripe phase compound La$_{1.48}$Nd$_{0.4}$Sr$_{0.12}$CuO$_4$. It was found that the intrinsic width in **Q**-space of the previously reported huge anomalous phonon softening and broadening is approximately 0.08r.l.u HWHM. A detailed comparison was also made to inelastic neutron scattering (INS) studies, which indicate a two-peak lineshape (with superimposed broad and narrow peaks) in the vicinity of the anomaly. The high resolution IXS data show that the narrow peak is mostly an artifact of the poor transverse **Q**-resolution of INS. Otherwise the agreement between the INS and IXS was excellent.

## I. Introduction

Recent INS measurements showed a strong anomaly in the Cu–O bond-stretching phonon in the copper oxide superconductors La$_{2-x}$Sr$_x$CuO$_4$ (with $x$=0.07, 0.15)[1-4], which was unexpected from conventional theory. The anomaly is strongest in La$_{1.875}$Ba$_{0.125}$CuO$_4$ and La$_{1.48}$Nd$_{0.4}$Sr$_{0.12}$CuO$_{4+\delta}$, compounds that exhibit spatially modulated charge and magnetic order, often called stripe order; it occurs at a wave vector corresponding to the charge order, **q**$_{co}$. The neutron data on the stripe phase show that the phonons of the bond-stretching branch dispersing from the zone center in the direction parallel to the Cu-O bond (the [1 0 0] direction) acquire a pronounced low energy tail at the reduced wavevector (0.25 0 0). Between this wavevector and the zone boundary (0.5 0 0) the phonon lineshapes may be fit with a two-peak structure. At the zone boundary the single peak lineshape is recovered. According to Refs. 3 and 4, the full spectrum of the bond-stretching vibration in La$_{1.875}$Ba$_{0.125}$CuO$_4$ may be fit with a superposition of two phonon branches of equal intensity as seen in fig 2a of Ref. 3. One branch has a monotonic, cosine-like downward dispersion in the [1 0 0] direction. In some sense, it can be considered as "normal" although it is well known that the corresponding branch is nearly flat in the undoped parent compound.[9] The other branch is highly anomalous in that it drops abruptly by more than 10 meV at about $\mathbf{q}_{co}$ = (0.25 0 0) and then rises back up towards the zone boundary to merge with the normal branch. The drop is much sharper than the instrumental resolution in both the longitudinal and the transverse directions, so the average over the wave vectors within the resolution ellipsoid produces the broad low-energy tail at $\mathbf{q}_{co}$.

Unfortunately, the INS experiments required focused beams, which reduced the **q**-resolution to about 15% of the Brilloiun zone, thus it has not been possible to determine the intrinsic **q**-width of the phonon anomaly either in the transverse or the longitudinal direction with sufficient precision. We decided to continue our measurements with inelastic x-ray scattering (IXS), which makes it possible to achieve a much higher wavevector resolution.

High resolution IXS is a relatively new technique, and only a few papers appeared on the bond stretching phonons in cuprates. Ref. 10, which focused on the electron-doped compound $Nd_{1.85}Ce_{0.15}CuO_4$ showed that the IXS phonon spectra are dominated by overlapping closely-spaced phonon peaks on top of a strong tail from the elastic line, which made it difficult to extract peak positions without assuming where they should be. A subsequent neutron scattering investigation performed when a large high quality sample was available produced data of superior quality, correcting some of the mode assignments.[11]

Uchiyama et al. obtained IXS data using a $0.2 \times 0.3 \times 0.3 mm^3$ $HgBa_2CuO_4$ sample, and showed that there is a minimum of the bond-stretching phonon dispersion in the 1 0 0 direction half-way to the zone boundary[12]. An investigation by Fukuda et al. demonstrated that of all the cuprates IXS works well for measuring the bond-stretching phonon in the $La_{2-x}Sr_xCuO_4$ family.[13] They clearly showed the anomalous phonon softening discussed above near the optimum doping level and demonstrated that phonon anomaly disappeared in overdoped compounds. However in both studies, the relatively poor energy resolution with Lorentzian tails, resulted not only in resolution broadening of the phonons, but also in a strong background due to the tail of the elastic diffuse scattering, which made it difficult to reliably extract the phonon linewidths. For the case of the $La_{2-x}Sr_xCuO_4$ family the availability of ~cc sized single crystals, has made it possible to obtain data with better statistical quality using INS[2,3,14].

The purpose of the current experiment was to perform measurements of the bond-stretching phonon in the LaSrCuO family that would enable us to extract the intrinsic width of the phonon anomaly in **q**. The idea was to use better energy resolution than in the previously published IXS studies (~1.7 meV FWHM) and to take advantage of the 2D analyzer array at BL35XU[15,16] at SPring-8 to measure transverse **q**-dependence of the phonon lineshape simultaneously with the longitudinal one. We focused on the stripe phase compound $La_{1.48}Nd_{0.4}Sr_{0.12}CuO_4$, because neutron scattering experiments indicate that the stripe phase has the narrowest intrinsic **q**-width of the phonon anomaly in both transverse and longitudinal directions. Furthermore, it was already thoroughly studied by neutrons, which made it possible to compare neutron and x-ray data on the same compound. We find that the two techniques produce virtually identical results with the main difference coming from different resolution functions. The half-width-at-half-maximum (HWHM) of the region in **q**-space where the anomalous phonon is found, is 0.08 r.l.u. in both the longitudinal and transverse directions. We also established that most of the intensity observed using INS in the "normal" branch discussed above is probably an artifact of poor wavevector resolution of the neutron measurements, although we cannot rule out an anomalous intrinsic lineshape of the phonon mode.

## II. Experimental Details

We used the inelastic x-ray scattering (IXS) spectrometer BL35XU at SPring-8[15] to precisely measure the width of the phonon anomaly in **q**-space in a high quality single crystal of $La_{1.48}Nd_{0.4}Sr_{0.12}CuO_4$. The sample was mounted in the standard configuration in the reflection mode. The mosaic measured on the (4 0 4) Bragg peak was 0.3º. The switch from INS to IXS was to improve both the precision and the wavevector (**Q**) – resolution, since the Q/energy resolution were a factor of 4/2 better respectively in our IXS measurements. (We note here that **Q** refers to the total wavevector transfer, whereas **q** refers to **Q** reduced to the first Brillouin zone.)

In order to improve the energy resolution and, just as importantly, to reduce background coming from the tail of the elastic line and low energy phonons, we used the (11 11 11) reflection of the Si monochromator and analyzer achieving 1.7 meV FWHM energy resolution. In this configuration the background was very small and came from the elastic line only and not from other phonons at the measured points. This configuration enabled us to isolate the bond-bending vibrations appearing near 59meV in the q-range of interest, from the bond-stretching phonons at higher energies.

The IXS instrument on BL35XU has a 2D array detector, with 3 rows of 4 analysers in each.[16] The sample was aligned with the **Q**-points along the [1 0 0] high symmetry direction measured by the top row of the detectors. The middle and bottom rows measured spectra with nonzero transverse components of **Q** (Fig. 1) . The size of the squares in figure 1 corresponds to the **Q**-resolution of the analyzers, which was 0.04 r.l.u. full width (FW). Slits limiting the analyzer acceptance were opened to allow maximum count rate: better **Q**-resolution is also possible (by a factor of ~5) in the event that one is not count-rate limited.

The measurements were performed at 7K on the $La_{1.48}Nd_{0.4}Sr_{0.12}CuO_4$ sample with the stoichiometry associated with stripe order.

## III. X-ray scattering results

Figure 2 shows the background-subtracted spectra that represent phonon scattering intensity. The background consisted of the tail of the elastic line due to the Lorentzian-like energy resolution function of the instrument and a contribution assumed to be energy independent due to both dark counts and stray x-rays and leaking into the detectors. The contribution from the elastic tail was obtained by first measuring plastic and then scaling together the intensities of plastic and our sample at the elastic position. This procedure may introduce some errors due to neglect of the tails from low-energy phonons and inelastic scattering from the plastic in the estimate of the true elastic tail, but we believe they should not exceed 20% of the reported signal. The exception is H=3.11, which is closer to the tail of the (3 0 0) Bragg peak. Here the elastic tail makes a significant contribution, but the phonon is very strong (due to a large structure factor) and narrow, so it was still possible to extract both the peak position and linewidth with high accuracy. Background rates varied from detector to detector between 0 and 0.01 on the scale in figure 2.

Previous neutron data showed that two longitudinal branches should appear in the investigated energy range. The bond-bending branch disperses upwards towards the zone boundary but stays mostly below 60meV. The bond-stretching branch disperses downwards in the [1 0 0] direction from the zone center energy of 85meV but stays mostly above 60meV.

The bond-stretching phonon anomaly at K=0 is clearly seen in the data. At Q=(3.11 0 0) the bond-stretching phonon is narrow and shows a completely "normal" behavior. At Q=(3.19 0 0) it already broadens and at Q=(3.27 0 0) it broadens even further. Here the low-energy tail of the stretching phonon overlaps with the weaker and narrower bond-bending phonon at 57meV. The fact that clear broadening is observed already at q=(0.19 0 0), means that there is an intrinsic **q**-width of the phonon anomaly in the longitudinal direction of about 0.08 r.l.u. HWHM.

We now discuss the evolution of the phonon lineshape in the transverse direction (K>0). The data are consistent with the previous claims of neutron studies that the strongest anomaly appears at the stripe-ordering wavevector and that it is narrow in the transverse direction.[3,4] The anomalous behavior is completely absent in all scans with K=0.16, whereas the scans with K=0.08 are intermediate between 0 and 0.16. The big difference with the neutron data is that the transverse q-resolution of our experiment was 0.04 r.l.u FW as opposed to 0.15 r.l.u. FWHM in the neutron experiments. Thus the above-mentioned partially developed phonon broadening at K=0.08 is intrinsic behavior, not a resolution effect, and we can conclude that the intrinsic q-width of the phonon anomaly in the transverse direction of about 0.08 r.l.u. HWHM, which is similar to the one observed in the longitudinal direction.

One seemingly puzzling feature is the apparent absence of the bond-bending mode in some of the scans at Q=(3.27 K 0) and Q=(3.35 K 0) where it should appear according to the shell model and previous neutron data. This is probably due to the fact that the energy region below 60 meV was counted half the time compared to the range between 65 and 75 meV. Thus the statistics there are relatively poor, and some scans show a more or less clear mode and others do not.

**IV. Comparison with neutron data.**

Figure 3 compares the neutron data for $La_{1.875}Ba_{0.125}CuO_4$ at **q**=(0.275 0 0)[1] and the x-ray data at **q**=(0.27 0 0) (a) and **q**=(0.27 0.08 0) (b) for $La_{1.48}Nd_{0.4}Sr_{0.12}CuO_4$ from the current experiment. (There are no data at this wavevector available for $La_{1.48}Nd_{0.4}Sr_{0.12}CuO_4$. A comparison with the neutron data on $La_{1.48}Nd_{0.4}Sr_{0.12}CuO_4$ at **q**=(0.25 0 0) gives similar results.) It is intriguing that the nominally K=0 neutron data are much closer to the K=0.08 x-ray data than to the K=0 x-ray data. As mentioned above, the neutron data are characterized by a two-peak lineshape with a relatively sharp higher energy "normal" peak combined with a broad lower-energy "anomalous" peak. In the x-ray data at K=0 (Fig. 3a) the "normal" peak is suppressed. We believe that this difference comes from the difference in resolution functions. Due to the poorer K-resolution, the neutron experiment picks up some intensity from narrow "normal" phonons at K≠0. This is probably the reason that the x-ray data taken with K=0.08 has the same lineshape as the nominally K=0 neutron data. On the other hand, the broad "anomalous" component observed by neutrons remains in the x-ray data. This observation shows that its previously reported large linewidth is intrinsic, and is not an artifact of a very steep dispersion.

Although both the neutron and x-ray data show that experimental anomalous phonon lineshapes are more complicated than simple Gaussians or Lorentzians, it is instructive to compare the results obtained by the two techniques on the same compound by fitting the phonons with single peak functions. The neutron/x-ray data were fit with the Gaussian/Lorentzian functions respectively, and the resolution-corrected results are shown in Fig. 4. It is clear, that despite different resolution functions as well as different background contributions, the fits give remarkably similar results. We believe that the differences around h=0.3 are due to the differences in the **Q**-resolutions of the two techniques.

V.     **Discussion**

A relatively simple model assuming stripe order,[17] can explain most of the observed features in the phonon measurements. According to this model, collective electronic excitations emanating from the stripe ordering wavevector interact with phonons at the intersection of the electronic and the phonon dispersion curves. The strongest coupling is expected for bond-stretching modes, because the charge modulation gives rise to a bond length modulation. Due to the steep dispersion of the electronic modes in both the transverse and the longitudinal directions, they interact with the phonons in a very limited range of reciprocal space. This model, however, predicts that the two branch behavior should be intrinsic, since stripes must introduce a mechanism of local breaking of the tetragonal symmetry of the $CuO_2$ plane distinguishing between the phonons propagating parallel and perpendicular to the stripes. Since the crystals are macroscopically twinned with respect to the stripe direction, each branch would contribute equally to the neutron spectra resulting in the normal and the anomalous branch. However, here we found that the previously reported two-branch behavior[3,4] is mostly a resolution effect with most intensity in the normal branch observed in INS coming from the nonzero K.

M. Vojta et al.[18] have shown that such symmetry breaking may occur only very locally when stripes are dynamic, since stripe fluctuations may occur in both directions simultaneously. The estimate of the **q**-width of the phonon anomaly allows us to extract the size of the coherent stripe domain in real space to be about 12x12 unit cells.

Alternatively, the phonon anomaly may be seen as a conventional Kohn anomaly due to Fermi surface nesting at $2\mathbf{k}_f$, where $\mathbf{k}_f$ is Fermi momentum, and not as a result of stripe order.[19] However, photoemission measurements[20] have shown that the phonon anomaly does not correspond to any Fermi surface nesting wavevector at $2\mathbf{k}_f$. A $4\mathbf{k}_f$ anomaly[21,22] may be consistent with short parallel segments of the Fermi surface observed by photoemission at the antinodal points. The relationship between the stripe order and the shape of the Fermi surface is not clear at the moment, so more work needs to be done to connect the phonon effect, the Fermi surface topology, and the stripes.

VI.    **Conclusions**

The agreement between the current x-ray and previous neutron results is remarkable if differences in resolution functions are taken into account. Thus our measurements definitively confirmed anomalous phonon broadening and softening previously observed by neutron scattering in the stripe phase of the cuprates.[3,4]

Most of the signal in the sharp peak on the high-energy side of the phonon lineshape in the neutron data[3] is an artifact of relatively poor transverse **q**-resolution of the neutron experiments, whereas the broad strongly renormalized component is intrinsic. However, even in the IXS data the bond-stretching phonon at q=(0.27 0 0) has a non-Lorentzian lineshape. It can be an artifact of poor statistics, or of the finite **q**-resolution, but at this point it is impossible to completely rule out a lineshape that is more complex than a simple Lorentzian peak.

The intrinsic **q**-width of the anomaly is around 0.08 r.l.u. HWHM both in the longitudinal and transverse directions. Assuming that the phonon anomaly results from the matching of the phonon and stripe propagation vectors, this width corresponds to a coherent stripe domain of 12x12 unit cells.

## VII. Acknowledgement

The authors would like to thank L. Pintschovius and J.M. Tranquada for helpful discussions and suggestions. D.R. was supported in part by International Frontier Center for Advanced Materials (IFCAM) at Tohoku University.

Figures:

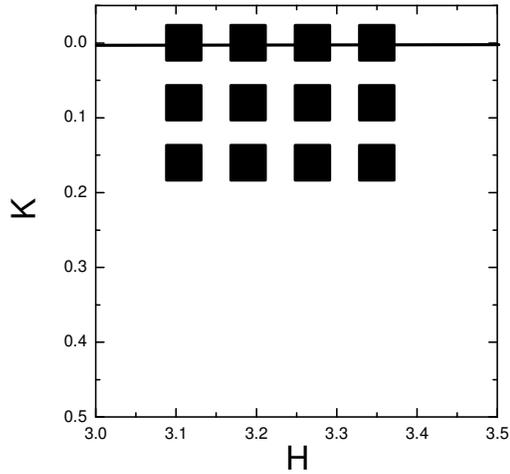

Figure 1. Schematic of the detector coverage of the 2D Brillouin zone of the Cu-O plane. The size of the squares corresponds to the experimental resolution. The top row of analyzers probes the high symmetry direction with $Q_x=0$. The other rows provide information on the nonzero transverse component of **Q**.

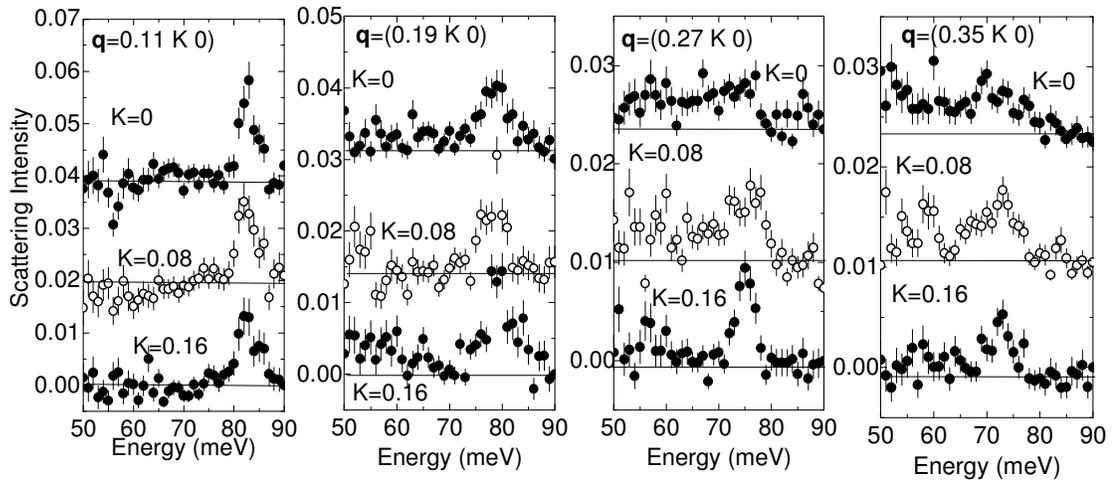

Figure 2. Energy scans after subtraction of the elastic tail as well as a constant term corresponding to the stray radiation as described in the text. The vertical scale is approximately counts per second, for an incident beam of some $4 \times 10^9$ photons/s onto the sample. Data was collected for approximately 2.5 days simultaneously into all spectra, with most of the time devoted to counting between 60 and 80 meV energy transfers.

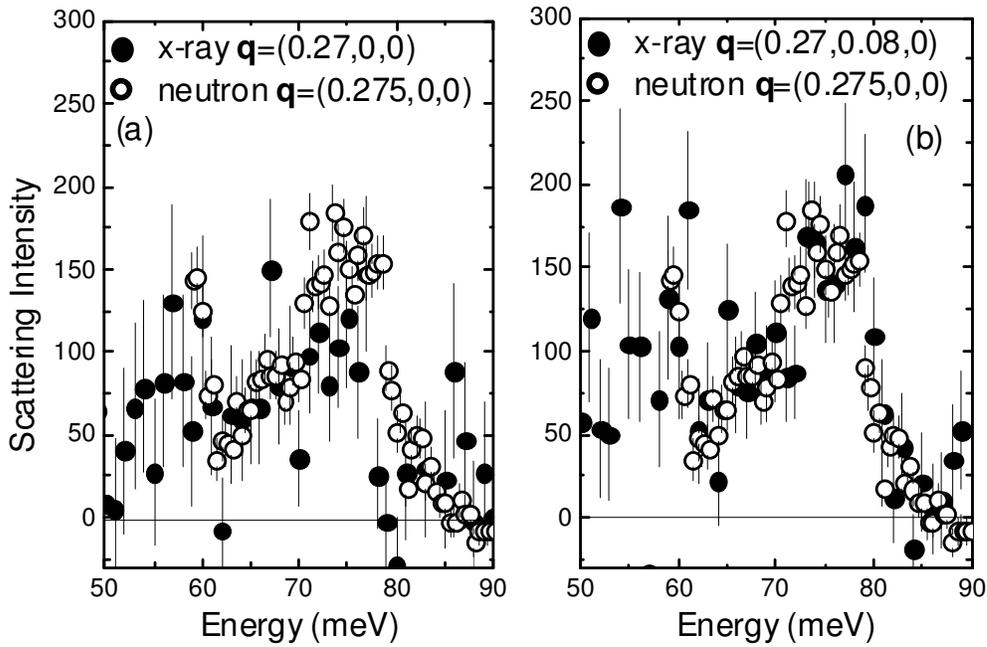

Figure 3. Comparison of neutron and x-ray spectra of $La_{1.48}Nd_{0.4}Sr_{0.12}CuO_4$ at $\mathbf{Q}=(3.27\ K\ 0)$, where K=0 (a) and K=0.08 (b) of the neutron spectrum of $La_{1.875}Ba_{0.125}CuO_4$ at $\mathbf{Q}=(4.725\ 0\ 0)^3$. The region between 60 and 80 meV in the x-ray data was counted much longer than others. The x-ray data in the two panels were scaled by the same factor.

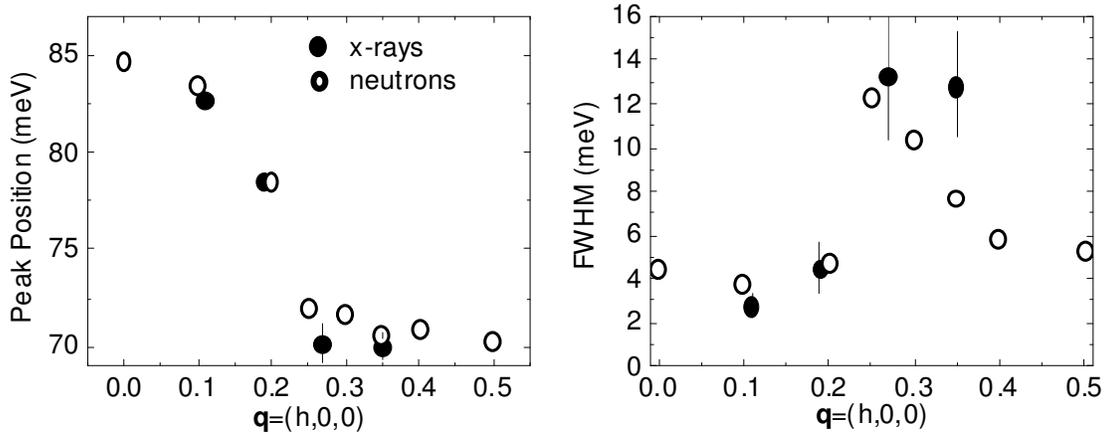

Figure 4 Comparison of the peak position and linewidth extracted from one-peak fits to the data above 60meV at K=0 measured on $La_{1.48}Nd_{0.4}Sr_{0.12}CuO_4$ by neutrons (Ref. 2) and x-rays (current study, see text).